\documentclass{article}

\usepackage[dblblindworkshop, final]{neurips_2025}
\usepackage[utf8]{inputenc} 
\usepackage[T1]{fontenc}    
\usepackage{hyperref}       
\usepackage{url}            
\usepackage{booktabs}       
\usepackage{amsfonts}       
\usepackage{amsmath}
\usepackage{nicefrac}       
\usepackage{microtype}      
\usepackage{xcolor}         

\usepackage{amsmath,amssymb,amsthm}
\usepackage{algorithm,algpseudocode}
\usepackage{graphicx}
\usepackage{xspace}
\usepackage{tikz-cd}
\usepackage{subcaption}
\usepackage{enumitem} 

\newcommand{\Collab}{{\sc CoLLAB}\xspace}
\newcommand{\ReCollab}{{\sc ReCoLLAB}\xspace}

\newcommand{\us}[1]{{#1}}

\title{\ReCollab: Retrieval-Augmented LLMs for Cooperative Ad-hoc Teammate Modeling}

\author{
  Conor Wallace, Umer Siddique, Yongcan Cao \\
  Department of Electrical and Computer Engineering, \\University of Texas at San Antonio \\
  \texttt{conor.wallace@my.utsa.edu}, \texttt{muhammadumer.siddique@my.utsa.edu}, \\
  \texttt{yongcan.cao@utsa.edu} \\
}

\begin{document}
\maketitle

\begin{abstract}
Ad-hoc teamwork (AHT) requires agents to infer the behavior of previously unseen teammates and adapt their policy accordingly. Conventional approaches often rely on fixed probabilistic models or classifiers, which can be brittle under partial observability and limited interaction. Large language models (LLMs) offer a flexible alternative: by mapping short behavioral traces into high-level hypotheses, they can serve as world models over teammate behavior. We introduce \Collab, a language-based framework that classifies partner types using a behavior rubric derived from trajectory features, and extend it to \ReCollab, which incorporates retrieval-augmented generation (RAG) to stabilize inference with exemplar trajectories. In the cooperative Overcooked environment, \Collab effectively distinguishes teammate types, while \ReCollab consistently improves adaptation across layouts, achieving Pareto-optimal trade-offs between classification accuracy and episodic return. These findings demonstrate the potential of LLMs as behavioral world models for AHT and highlight the importance of retrieval grounding in challenging coordination settings.
\end{abstract}

\section{Introduction}
\label{sec:intro}

Multi-agent reinforcement learning (MARL) has achieved remarkable success in several domains, from zero-sum games such as Go~\citep{silver2016mastering}, to robotics~\citep{de2020deep}, autonomous driving~\citep{zhou2020smarts}, and cooperative control tasks~\citep{samvelyan2019starcraft,lin2023tizero}. 
Within MARL, cooperative MARL focuses on training teams of agents to solve a common task by interacting with the environment and with each other. 
Although this particular setting has shown impressive results in controlled environments~\citep{rashid2020monotonic,rashid2020weighted,son2019qtran,yu2022surprising}, it typically assumes that all agents are trained under the same algorithm, which limits its applicability in heterogeneous and realistic settings. Moreover, despite advances in addressing challenges such as non-stationarity~\citep{nekoei2023dealing}, credit assignment~\citep{zhou2020learning}, cooperative MARL requires that all teammates be known in advance. 
However, in practice, many real-world settings often involve heterogeneous agents that are not jointly trained in a shared environment, such as fleets of UAVs with different specifications working together, or different autonomous vehicles sharing the same roads.

To address such scenarios, the \emph{ad-hoc teamwork} (AHT) framework was proposed~\citep{stone2010ad}, which aims at enabling an autonomous agent to collaborate effectively with previously unseen teammates without prior coordination.
Successful AHT requires the ability to rapidly infer the behavior of a teammate and adapt one's own strategy accordingly. This capability is especially important in MARL, where partner policies can be diverse and partially observable. A central challenge in this setting is \emph{teammate modeling}, which refers to constructing representations or beliefs about another agent’s latent policy type based on partial observations of its behavior~\citep{albrecht2018autonomous}. Prior approaches include probabilistic reasoning over discrete teammate types, as in PLASTIC~\citep{barrett2017plastic}, and contract-based frameworks such as M\textsuperscript{3}RL~\citep{shu2019m3rl}, which employ managerial agents to guide self-interested workers. Model-based approaches such as TEAMSTER~\citep{ribeiro2023teamster} further propose separate modeling of the environment and teammate behaviors in model-based ad hoc teamwork. Meanwhile, recent work on the $N$-Agent Ad Hoc Teamwork (NAHT) framework~\citep{wang2024n} extends AHT to dynamic settings with varying teammates. While effective in some domains, these methods require carefully designed state abstractions and likelihood models that may be brittle under partial observability or in more semantically complex environments.

Large Language Models (LLMs) offer a complementary approach by reasoning over natural-language summaries of observed behavior to infer teammate types \citep{gao2024large}. LLMs can consume rich, structured prompts that describe recent interaction histories and generate semantically grounded inferences about teammate behavior. Although recent works have demonstrated LLM-based reasoning in competitive games~\citep{meta_diplomacy, richelieu2024self} and embodied decision-making~\citep{li2024embodied}, their application as \emph{structured world models over teammates} in cooperative MARL domains remains underexplored. Moreover, retrieval-augmented generation (RAG)~\citep{rag_survey} provides a mechanism for grounding LLM predictions in prior experience by retrieving relevant examples from offline trajectories, which can potentially enhance the robustness of LLM-based agents in ambiguous settings.

In this paper, we fill this gap by proposing a novel framework that leverages LLMs as world models for AHT. Specifically, we propose COoperative LLm-based Agent Belief or \Collab, which classifies the type of an unknown teammate based on their recent trajectory history and routes the interaction to a pre-trained best-response policy. Building on recent advances in retrieval augmentation, we further introduce \ReCollab, a retrieval-augmented variant that enriches the LLM prompt with retrieved summaries of similar teammate behaviors from a labeled trajectory database. To support experimental evaluation, we also contribute a labeled dataset for the cooperative Overcooked domain~\citep{carroll2019utility, jaxmarl2024}, which contains five different teammate behavior types induced via reward shaping.


Our main contributions are as follows:
\begin{enumerate}[left=1.5pt, nosep]
    \item \textbf{Behavioral World Modeling:} We formulate LLM-based teammate type classification as a form of world modeling for AHT.\\

    \item \textbf{Teammate Reasoning with Retrieval:} We propose \Collab and \ReCollab, bridging prompt-based reasoning with retrieval-augmented grounding in MARL.\\
    
    \item \textbf{Robust Early Teammate Identification:} We empirically demonstrate their effectiveness in the Overcooked environment and evaluate our methods against established baselines. Our results demonstrate that LLM-based teammate world models achieve competitive or superior early-type identification compared to baselines, and that retrieval augmentation improves robustness in ambiguous or noisy settings. 
\end{enumerate}

\section{Related Work}
\label{sec:related}

\paragraph{Ad-hoc Teamwork.}
Ad-hoc teamwork (AHT) is a long-standing problem in multi-agent systems, which focuses on enabling autonomous agents to collaborate effectively with previously unseen teammates without prior coordination or shared knowledge~\citep{stone2010ad}.
Early approaches focused on explicit \emph{teammate modeling}, where the agent maintains a belief over possible teammate types and adapts its policy accordingly. ~\citet{barrett2013teamwork} explore robust teaming under limited information, and demonstrated applications to more complex domains such as robot soccer~\citep{barrett2015cooperating}. Building on this, the PLASTIC framework~\citep{barrett2017plastic} represents a canonical example in the teammate modeling literature. It uses Bayesian belief updates over a discrete teammate-type space combined with best-response policy selection. Extensions include online learning of teammate models~\citep{albrecht2018autonomous}, policy reuse, and intention recognition. Model-based methods such as TEAMSTER~\citep{ribeiro2023teamster} further decouple environment and teammate modeling, while the $N$-Agent Ad Hoc Teamwork (NAHT) framework~\citep{wang2024n} extends AHT to settings with dynamically varying teammates. While these approaches yield strong performance, they typically rely on carefully designed state abstractions or likelihood models, which may be brittle in partially observable or semantically complex environments.

\paragraph{Multi-Agent Reinforcement Learning.}
Multi-agent reinforcement learning (MARL) extends single-agent RL to settings where multiple agents interact within a shared environment. 
In MARL, most of the work focuses on value decomposition, including VDN~\citep{sunehag2017value} and QMIX~\citep{rashid2020monotonic}, which factorize joint action-value functions into agent-wise utilities, and QTRAN~\citep{son2019qtran}, which generalizes decomposition via additional value function constraints.
In policy gradient methods, Independent PPO (IPPO)~\citep{de2020independent} and Multi-Agent PPO (MAPPO)~\citep{yu2022surprising} train agents with decentralized policies using independent or centralized critics, respectively. 
In contrast to these MARL methods, which assume training of all agents, AHT considers a single agent that must collaborate with unknown teammates in control of their own actions. 
In this paper, we adopt IPPO as our base MARL algorithm.
For empirical evaluation, we use the cooperative Overcooked environment~\citep{carroll2019utility}, a benchmark requiring coordination and division of labor, implemented in the JaxMARL framework~\citep{jaxmarl2024}. 
This two-player setting enables us to induce distinct teammate types for evaluating our proposed \Collab and \ReCollab methods.

\paragraph{LLMs for MARL.}
LLMs have recently been integrated with multi-agent decision making to enable agents that reason about partner intentions and coordinate through language. In the strategic game of Diplomacy, \citet{meta_diplomacy} combined LLM-based communication with search-based planning to achieve human-level play. In cooperative settings, ProAgent shows that LLMs can act as proactive teammates: it uses language to infer task context, anticipate partner needs, and adapt policies for zero-shot coordination, yielding strong gains across cooperative benchmarks \citep{zhang2024proagent}. A recent study on Mutual Theory of Mind (ToM) deploys an LLM agent with ToM and communication modules in a real-time shared-workspace task, finding that language-based modeling improves mutual understanding even when raw task performance gains are modest—underscoring both the promise and limits of LLMs for behavior inference in interactive collaboration \citep{zhang2024mutual}.


\paragraph{Retrieval-Augmented Generation.}
Retrieval-Augmented Generation (RAG)~\citep{rag_survey} combines parametric LLM knowledge with non-parametric retrieval from a database of relevant examples. RAG  has been widely used in knowledge-intensive NLP tasks and has recently been applied to embodied agents and multi-modal reasoning~\citep{visualrag2025}. In MARL, retrieval has been explored for retrieving relevant past multi-agent behaviors from a skill database to augment limited demonstrations, enabling more effective policy learning for cooperative mobile robot manipulation tasks~\citep{kuroki2024iros}. However, its integration with LLM-based teammate modeling has not yet been studied. Our proposed \ReCollab applies RAG to retrieve offline trajectory snippets to ground LLM predictions in concrete prior experience, which further improves robustness in AHT.


\begin{figure*}[t!]
\centering
\includegraphics[width=1.0\linewidth]{Images/recollab.pdf}
\caption{\textbf{System diagram of \Collab and \ReCollab.} The Overcooked environment produces observed trajectories from a controlled agent interacting with a teammate. These trajectories are transformed into prototype feature vectors (e.g., action histograms, dwell times, cumulative reward), which are matched against behavior rubrics to model teammate types. In \Collab, an LLM classifies the teammate type directly from the behavior rubric. In \ReCollab, the LLM additionally conditions on retrieved exemplar trajectories from a database, grounding its classification. Both approaches output a predicted teammate type with associated confidence and rationale, which is used to select the best-response policy from a library of trained policies.}
\label{fig:pipeline}
\end{figure*}

\section{\us{Background and }Problem Formulation}
\label{sec:problem}

We model the cooperative ad-hoc teamwork scenario as a two-agent partially observable Markov game (POMG) \us{$\mathcal{G} = \left\langle \mathcal{S}, \mathcal{A}_0, \mathcal{A}_1,  \Omega_0, \Omega_1, P, r_0, r_1, \gamma \right\rangle$
where $\mathcal{S}$ is the set of environment states, $\mathcal{A}_0$ and $\mathcal{A}_1$ are the discrete action spaces for the \emph{teammate} and the \emph{controlled agent}, respectively. In this model, $P: \mathcal{S} \times \mathcal{A}_0 \times \mathcal{A}_1 \to \Delta(\mathcal{S})$ defines the state transition function,  $\Omega_i:\mathcal{S}\to \Delta(\mathcal{O}_i)$ is the observation function for agent $i\in\{0,1\}$ producing partial observations $o_t^i \sim \Omega_i(s_t)$, rewards are $r_i:\mathcal{S}\times \mathcal{A}_0 \times \mathcal{A}_1 \to \mathbb{R}$, and discount factor is $\gamma \in (0,1]$.
Since we consider the cooperative setting, we assume $r_0 =r_1=r$.} 

\us{In this model, agents do not have access to the true transition function $P$ (and may not know the exact teammate policy), and must act based on partial observations. The goal of the agent $i$ is to learn a policy $\pi$ (possibly history dependent), which is a mapping $\pi^i:\mathcal{H}^i \to \Delta(\mathcal{A}_i)$, where $\mathcal{H}^i$ denotes the space of local action-observation histories for agent $i$. Unless otherwise stated, we consider stochastic Markov policies for the controlled agent and the teammate may exhibit diverse behaviors captured by the considered model type. In this mode, the teammate's policy $\pi^{0,\tau}$ which is unknown to the controlled agent and is drawn (per trajectory)} from a finite set of $M$ possible teammate types $\mathcal{T} = \{\tau_1, \tau_2, \dots, \tau_M\}$ where each type represents a distinct behavioral strategy, potentially induced by different training curricula or reward shaping. The controlled agent has access to a corresponding set of best-response policies: $\Pi = \{\pi^{1,\tau_1}, \pi^{1,\tau_2}, \dots, \pi^{1,\tau_M}\}$ where $\pi^{1,\tau_m}$ is optimized to maximize expected return when paired with teammate of type $\tau_m$.


\us{\paragraph{Problem statement.}
We consider the cooperative ad-hoc teamwork scenario where the goal is to infer the teammate type $\tau$ as quickly and accurately as possible from the changing history $h_t$ and select actions using the corresponding best-response policy $\pi^{1, \hat{\tau}}$ to maximize the team's discounted return.  We formalize this problem of \textbf{teammate type classification} as learning a mapping function: 
\begin{align}\label{eq:problem}
    f_\theta:\mathcal{H}\to \mathcal{T}, \quad \hat{\tau}_t \;=\; f_\theta(h_t),
\end{align}
where $\mathcal{H}$ is the space of possible histories.  In this paper, we model  $f_\theta$ as either a prompt-based LLM (\Collab) or a retrieval-augmented LLM (\ReCollab), where $h_t$ is supplemented with retrieved summaries from an offline trajectory database to improve robustness.}

\section{Methods}
\label{sec:methods}
\us{To solve the problem of teammate type classification (\ref{eq:problem}), we frame this as a world modeling problem wherein the controlled agent, which is restricted from a fundamental source of information, must infer the teammate's behavior type $\tau \in \mathcal{T}$ from partial observations in order to form the best-response policy. To achieve this, we leverage LLMs as lightweight world models that reason over structured descriptions of teammate behavior. We propose two methods: \Collab, which classifies using statistical prototypes summarized in a behavior rubric, and \ReCollab, which augments \Collab with retrieval from an offline trajectory database to resolve ambiguities and enhance robustness.}

\subsection{Behavior Modeling \us{and Rubric Construction}}
{In complex multi-agent environments such as Overcooked~\citep{carroll2019utility}, raw trajectories are typically high-dimensional and thus difficult for LLMs to interpret directly. This occurs for two main reasons: 1) LLMs struggle to extract relevant features from episodic trajectories because they lack an understanding of the underlying game mechanics, and 2) they lack knowledge of the specific behavioral characteristics of each teammate type. To address these limitations, we construct a higher-level rubric of teammate behaviors from an early probing window that captures the most relevant features required to discriminate between teammate types and is easy for an LLM to ingest. Formally, let $\mathbf{f} = (f_1, f_2, \ldots, f_m)$ denote the feature vector computed over the first $P$ steps of an episode probing period, where each feature $f_j$ encodes a behavioral statistic such as dwell time near
a station, number of interactions with pots, or cumulative reward.

\us{\paragraph{Feature selection.}
To retain only discriminative features, we estimate the
mutual information (MI), \cite{shannon1948}, between each feature $f_j$ and the teammate type label $\tau \in \mathcal{T}$:}
\[
I(f_j; \tau) = \sum_{f_j, \tau} p(f_j, \tau) \log \frac{p(f_j, \tau)}{p(f_j)\,p(\tau)}.
\]
This ranking highlights which behavioral dimensions most strongly
reduce uncertainty about $\tau$. Empirically, features such as dwell time
at the window or plate pile exhibit high MI, whereas handoff events
and blocked events are less informative. We keep the top-$r$ ranked
features (typically $r \leq 20$) to define the rubric.

\paragraph{Rubric construction.}
For each teammate type $\tau \in \mathcal{T}$, we compute summary
statistics (mean and standard deviation) of the selected features
across multiple offline episodes:
\[
\mu_{j,\tau} = \mathbb{E}[f_j \mid \tau], 
\quad \sigma_{j,\tau} = \sqrt{\text{Var}[f_j \mid \tau]}.
\]
The rubric function $r(\mathcal{T}) = \texttt{Rubric}(\mathcal{T})$ thus consists of behavior prototypes $\{ (\mu_{j,\tau}, \sigma_{j,\tau}) : j=1,\ldots,r\}$ for each type $\tau \in \mathcal{T}$. These serve as reference points describing in natural language how a typical teammate allocates its time and actions in the probe window.

\subsection{\Collab}
\us{We first introduce \Collab, our base method for teammate type
classification in AHT. \Collab leverages an LLM as an implicit world model over teammates and classifies the teammate by comparing observed behavior fingerprints against rubric prototypes. The observed behavior prototypes
$\mathbf{f} = (f_1, \ldots, f_r)$ from the probe phase are converted
into a structured natural language description, $d(\textbf{f}) = \texttt{Describe}(\textbf{f})$, which is presented to the LLM along with the behavior rubric. The model outputs the predicted type as:}
\[
    \hat{\tau} = f_{\theta}( d(\textbf{f}), r(\mathcal{T}) \big),
\]
where $f_{\theta}$ denotes the LLM, $d(\textbf{f})$ is the language-based representation of the prototype features, and $r(\mathcal{T})$ is the behavioral rubric to follow. \us{\Collab exploits the LLM's reasoning ability to interpret structured cues without training a task-specific classifier.}

\us{\subsection{\ReCollab}
Although \Collab classifies teammates by comparing observed fingerprints against rubric prototypes, it can fail miserably in settings when multiple teammate types yield overlapping statistics (e.g., in the Overcooked environment \texttt{Plate}
vs. \texttt{Mixed}). To address this issue, we introduce \ReCollab, a
retrieval-augmented variant that grounds LLM reasoning in concrete
trajectory exemplars.}

We build an offline database $\mathcal{D} = \{(h_P^{(i)}, \tau) \}_{i=1}^N$ from rollouts up until the probing time $P$ of controlled agents paired with each teammate type $\tau \in \mathcal{T}$. We collect $N$ probing trajectories for each teammate type, each of which utilizes a different random seed to ensure a diverse search database. The behavior feature vector $\textbf{f}^{(i)}$ is computed from trajectory $h_P^{(i)}$, which is then converted to a natural language description using $d(\textbf{f}^{(i)}) = \texttt{Describe}(\textbf{f}^{(i)})$. This is then embedded into a vector space using an encoder $E_P = f_\phi(d(\textbf{f}^{(i)}))$, where $f_{\phi}$ is a language embedding model. These embeddings form the keys for similarity search. \us{At inference, given an observed probe trajectory $h_P$ and its corresponding behavior prototype description $d(\textbf{f})$, we compute its embedding $E_P$ and retrieve the top-$k$ nearest neighbors from $\mathcal{D}$:
\[
\mathcal{R}(d(\textbf{f})) =
\{(d(\textbf{f}), d(\textbf{f}^{(i)})) : i \in \text{TopK}_k \big[ s(\phi(d(\textbf{f})), \phi(d(\textbf{f}^{(i)}))) \big] \},
\]
where $s(\cdot,\cdot)$ is a cosine similarity function. 
The LLM is then prompted with the structured description of the observed fingerprint $d(\mathbf{f})$, the rubric $r(\mathcal{T})$, and the retrieved exemplars $\mathcal{R}(d(\textbf{f}))$. This enhances the prompt with grounded evidence of how real teammates of each type behave under similar probe conditions and outputs:}
\[
    \hat{\tau} = f_{\theta}( d(\textbf{f}), r(\mathcal{T}), \mathcal{R}(d(\textbf{f})) \big).
\]

\us{By grounding abstract rubric features with retrieval, \ReCollab mitigates classification errors in ambiguous settings and stabilizes predictions. }


\subsection{Policy Adaptation}
\paragraph{Policy adaptation.}
Once a type $\hat{\tau}$ is predicted, the controlled agent selects the
corresponding best-response policy $\pi^{0,\hat{\tau}}$ from the policy
library $\Pi$. To prevent instability from repeated switching, we route policies only once, immediately after the probe window. This means that a major hyperparameter in our framework is the probe length $P$ of this early trajectory window. The longer this period, the more stable the prediction will be, but at the cost of potential gains in cumulative reward. Both \Collab and \ReCollab integrate information-theoretic fingerprints with language-model reasoning to serve as lightweight behavior models.
Rather than fitting a parametric classifier, our methods exploit the
LLM’s ability to interpret structured cues against a rubric of known
behavioral prototypes, functioning as a world model that infers latent
teammate types during the probe phase.

\begin{table}[t]
\centering
\caption{Classification accuracy across three Overcooked layouts (mean$\pm$std over 5 seeds).}
\label{tab:classification}
\setlength{\tabcolsep}{5pt}
\begin{tabular}{lccc}
\toprule
Method & Cramped Room & Asymmetric Advantage & Coordination Ring \\
\midrule
Random              & 0.20{\tiny$\pm$}0.01 & 0.20{\tiny$\pm$}0.01 & 0.20{\tiny$\pm$}0.01 \\
Logistic Regression & \textbf{0.96}{\tiny$\pm$}0.00 & 0.69{\tiny$\pm$}0.15 & 0.81{\tiny$\pm$}0.14 \\
PLASTIC             & 0.81{\tiny$\pm$}0.08 & 0.69{\tiny$\pm$}0.15 & 0.58{\tiny$\pm$}0.17 \\
\Collab              & 0.66{\tiny$\pm$}0.19 & 0.39{\tiny$\pm$}0.00 & 0.35{\tiny$\pm$}0.14 \\
\ReCollab ($k=5$)      & 0.92{\tiny$\pm$}0.08 & \textbf{0.77}{\tiny$\pm$}0.12 & \textbf{0.96}{\tiny$\pm$}0.00 \\
\bottomrule
\end{tabular}
\end{table}

\begin{table}[t]
\centering
\caption{Cumulative returns across three Overcooked layouts (mean$\pm$std over 5 seeds).}
\label{tab:returns}
\setlength{\tabcolsep}{5pt}
\begin{tabular}{lccc}
\toprule
Method & Cramped Room & Asymmetric Advantage & Coordination Ring \\
\midrule
Oracle              & 188.0{\tiny$\pm$}1.6 & 272.0{\tiny$\pm$}51.5 & 188.0{\tiny$\pm$}32.5 \\
\midrule
Static              & 45.6{\tiny$\pm$}3.2 & 189.6{\tiny$\pm$}7.4 & 44.0{\tiny$\pm$}0.00 \\
Random              & 58.4{\tiny$\pm$}17.8 & 116.0{\tiny$\pm$}14.3 & 86.4{\tiny$\pm$}20.3 \\
Logistic Regression & \textbf{129.6}{\tiny$\pm$}10.9 & \textbf{200.8}{\tiny$\pm$}12.0 & 140.0{\tiny$\pm$}41.4 \\
PLASTIC             & 119.2{\tiny$\pm$}21.1 & 200.8{\tiny$\pm$}12.0 & 133.6{\tiny$\pm$}39.6 \\
CoLLAB              & 103.2{\tiny$\pm$}23.7 & 149.6{\tiny$\pm$}14.2 & 66.4{\tiny$\pm$}21.0 \\
ReCoLLAB ($k=5$)      & 120.8{\tiny$\pm$}15.7 & 181.6{\tiny$\pm$}22.1 & \textbf{146.4}{\tiny$\pm$}36.5 \\
\bottomrule
\end{tabular}
\end{table}

\section{Experimental Setup and Metrics}
\label{sec:experiments}

\us{\paragraph{Environments and Layouts.}
We evaluate all methods in the cooperative Overcooked environment using the JaxMARL framework~\citep{jaxmarl2024}. To capture diverse coordination challenges, we select three layouts with distinct structural properties. Our first layout \textbf{Cramped Room}, is a small, narrow kitchen where agents must share a tight space to access ingredients, pots, plates, and the serving window, making blocking inevitable and coordination essential. Our second layout from Overcooked is \textbf{Asymmetric Advantage}, which is an asymmetric kitchen layout where one agent has direct access to key stations (e.g., onion and pot), while the other must navigate a longer path, creating strong role specialization pressures. Our last considered environment is \textbf{Coordination Ring}, which is a circular kitchen where onions, pots, plates, and the serving window are arranged sequentially around a loop, requiring agents to coordinate their direction of movement to avoid blocking. These environments present increasing levels of difficulty, ranging from relatively constrained coordination to highly structured role specialization.}

\paragraph{Teammates and Best Response Policies.}
\us{Episodes are $400$ timesteps long, with synchronous actions. We train five fixed teammate policies $\pi^{0, \tau}$, one for each behavior type induced via reward shaping. These include \emph{default} (no reward shaping), \emph{pot-focused} (reward for placing onions in the pot), \emph{plate-focused} (reward for picking up plates), \emph{serve-focused} (reward for delivering completed soups), and \emph{mixed} (combined shaping to encourage proficiency across subtasks).} Each teammate policy is trained with PPO for 6 million timesteps. For each type $\tau \in \mathcal{T}$, the controlled agent policy $\pi^{1, \tau}$ is trained with standard rewards while paired with $\pi^{0, \tau}$, yeilding a set of best-response policies.

\paragraph{Offline Dataset for RAG.}
We collect a dataset of $10$ evaluation episodes per teammate type per layout. During the probing phase of length $P=20$ timesteps, we always use the best-response policy to the \emph{default} teammate. Once the probe trajectory has been converted to natural language, we embed the trajectory using the \texttt{text-embedding-3-large} language embedding model from OpenAI \citep{openai-text-embedding-3-large}. This dataset is used exclusively for retrieval in \ReCollab and is disjoint from evaluation episodes.

\paragraph{Evaluation and Metrics.}
At the beginning of every episode, we sample a teammate type $\tau \sim \mathcal{T}$. As described in the previous section, the best response policy to the \emph{default} teammate is used during the probing phase. We report the teammate type classification as well as the cumulative reward over an episode. For both \Collab and \ReCollab, we utilize \texttt{GPT-5} as the base LLM \citep{openai-gpt5}.

\us{\paragraph{Baselines.}
We compare \Collab and \ReCollab against several baselines. These include \textbf{Oracle}, which has access to the ground-truth teammate type at $t=0$, \textbf{Static}, which maintains the same default policy throughout the episode, and \textbf{PLASTIC}~\citep{barrett2017plastic}  baseline, which performs Bayesian belief update with handcrafted likelihood functions over teammate actions. As a sanity check,  we also compare our methods with the \textbf{Random} baseline. This baseline randomly chooses the best-response policy at every time step $t$. Finally, the \textbf{Logistic Regression}  baseline performs the logistic regression classifier fit to the features from the behavior rubric $r(\mathcal{T})$.}  


\begin{figure}
    \centering
    \begin{subfigure}[b]{0.33\textwidth}
        \includegraphics[width=\linewidth]{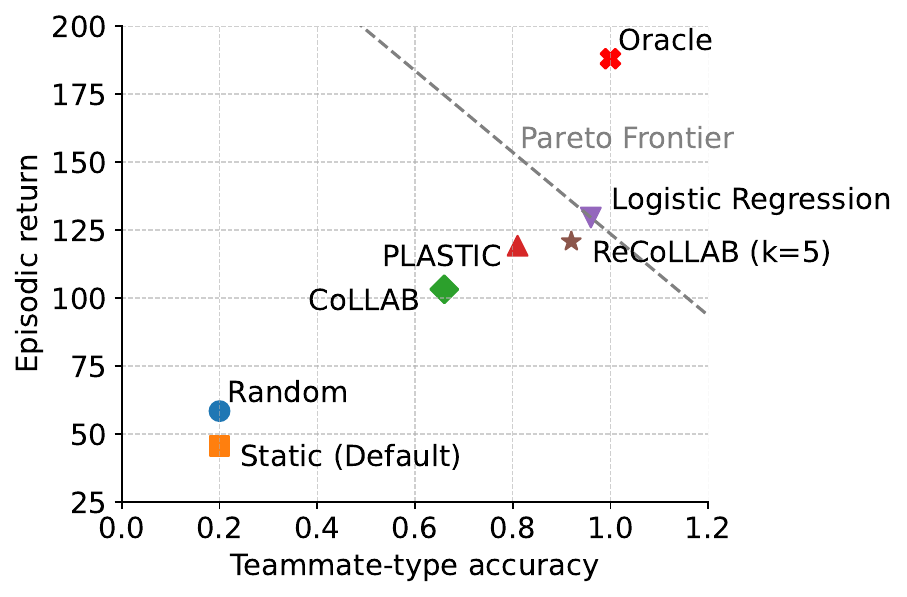}
        \caption{Cramped Room}
        \label{fig:pareto_sub1}
    \end{subfigure}\hfill
    \begin{subfigure}[b]{0.33\textwidth}
        \includegraphics[width=\linewidth]{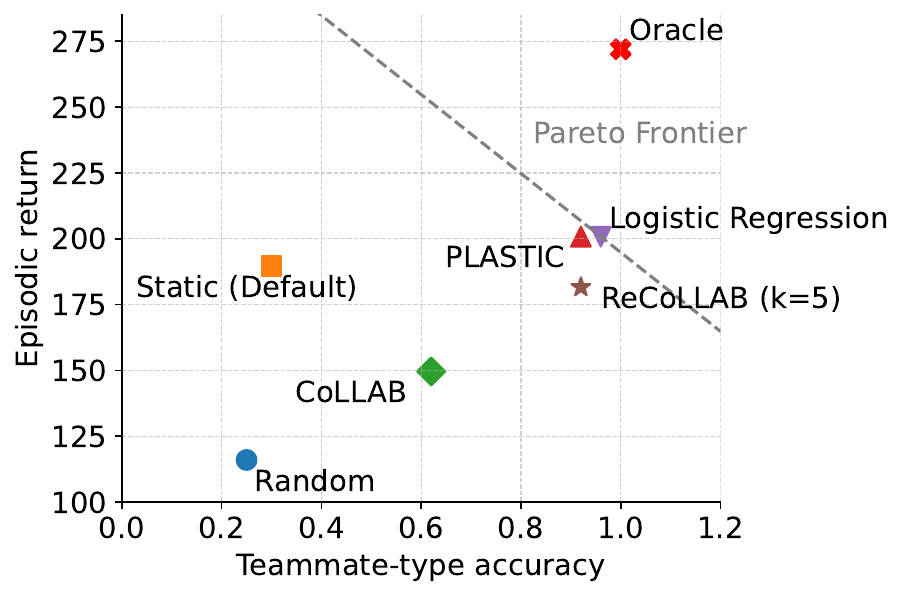}
        \caption{Asymmetric Advantages}
        \label{fig:pareto_sub2}
    \end{subfigure}
    \begin{subfigure}[b]{0.33\textwidth}
        \includegraphics[width=\linewidth]{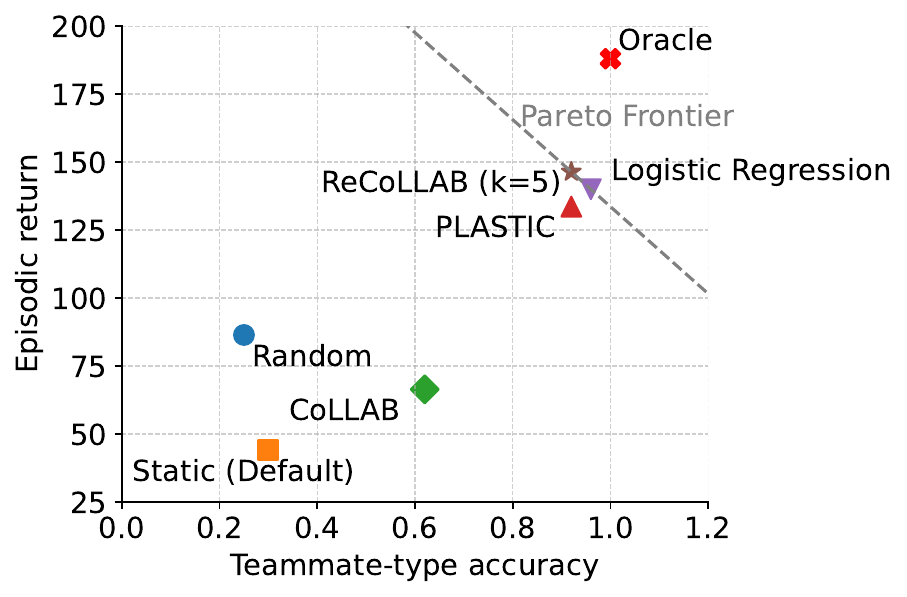}
        \caption{Coordination Ring}
        \label{fig:pareto_sub3}
    \end{subfigure}
    \caption{\textbf{Pareto Frontier Study.} We plot the teammate-type classification accuracy vs. the obtained cumulative reward and estimate the Pareto optimal frontier. \ReCollab consistently lies near or directly on the Pareto frontier.}
    \label{fig:pareto}
\end{figure}

\section{Results and Discussion}

\paragraph{Teammate type classification.} 
Table \ref{tab:classification} reports teammate-type classification accuracy across three Overcooked layouts. We observe that a simple Logistic Regression baseline achieves surprisingly strong performance, attaining the highest accuracy in \emph{Cramped Room} and remaining competitive in both \emph{Asymmetric Advantage} and \emph{Coordination Ring}. This finding highlights that the engineered fingerprint features already provide significant discriminative power, such that a lightweight linear classifier can separate teammate types effectively. By contrast, PLASTIC displays variable performance: it achieves strong results in \emph{Cramped Room} but struggles in the other two layouts, reflecting the brittleness of Bayesian updating when teammate behaviors are overlapping or sparsely expressed. \Collab, which relies only on rubric-based prompting of an LLM, consistently underperforms across layouts. This suggests that the rubric alone provides insufficient grounding for LLM classification. In contrast, \ReCollab substantially improves robustness by incorporating retrieval, outperforming PLASTIC in all three layouts. These results indicate that retrieval grounding stabilizes the LLM’s performance, particularly in more challenging layouts where teammate behaviors are less easily separable.

\paragraph{Cumulative returns.} 
Table~\ref{tab:returns} compares the cumulative rewards achieved under each method when the controlled agent adapts its best-response policy based on the predicted teammate type. As expected, the Oracle establishes the performance ceiling, while random and static policies tend to stand as the performance floor. Among adaptive methods, Logistic Regression, PLASTIC, and \ReCollab all produce substantially higher returns than \Collab, confirming the importance of accurate teammate modeling for ad-hoc teamwork. Notably, \ReCollab edges out PLASTIC in \emph{Coordination Ring} and surpasses it as well as Logistic Regression in \emph{Coordination Ring}. Logistic Regression again performs very competitively, particularly in \emph{Asymmetric Advantage}, suggesting that in certain layouts simple classifiers suffice for effective adaptation. 

\paragraph{Discussion.} 
Figure \ref{fig:pareto} plots each method's cumulative reward and teammate-type classification accuracy against each other and estimates the Pareto optimal frontier. \ReCollab consistently lies near, or directly on this frontier. Overall, the results reveal three main insights. First, the strong performance of Logistic Regression highlights the surprising discriminative power of fingerprint features, reinforcing the need to compare against lightweight baselines. Second, \ReCollab consistently improves upon \Collab and in many cases rivals or surpasses PLASTIC, especially in more difficult layouts. This demonstrates that retrieval grounding provides stability and robustness beyond what rubric-based prompting alone can achieve. Third, the divergence between classification accuracy and cumulative return underscores that accurate teammate-type inference is necessary but not sufficient; robustness to misclassification and stability of policy switching also play critical roles in maximizing reward. Taken together, these findings suggest that retrieval-grounded LLMs offer an interpretable and extensible alternative to Bayesian methods, while also revealing key limitations of current trajectory representations and prompting strategies. We view the development of richer fingerprint features, hybridized approaches combining retrieval with probabilistic priors, and scaling to more diverse teammate behaviors as promising directions for future work.

\begin{figure}
    \centering
    \begin{subfigure}[b]{0.47\textwidth}
        \includegraphics[width=\linewidth]{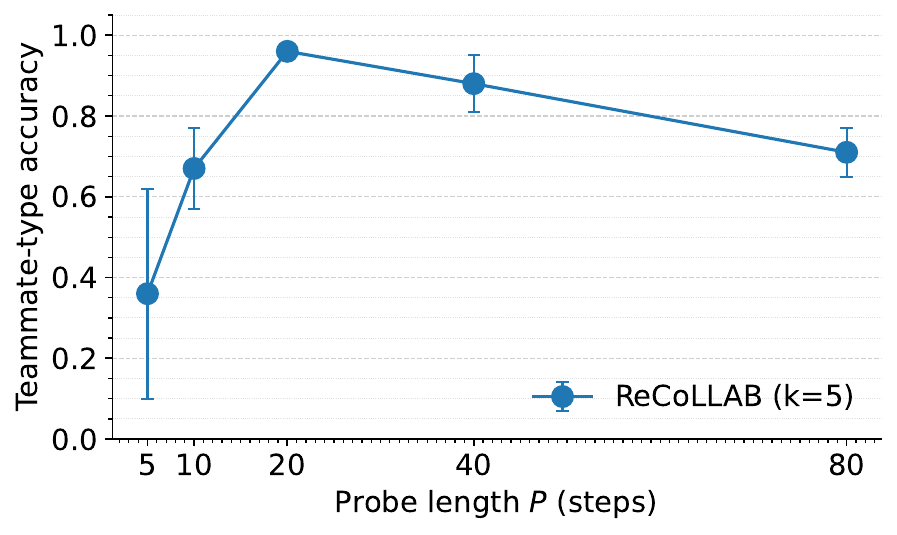}
        \caption{Classification Accuracy vs. Probe-length}
        \label{fig:probe_sub1}
    \end{subfigure}\hfill
    \begin{subfigure}[b]{0.47\textwidth}
        \includegraphics[width=\linewidth]{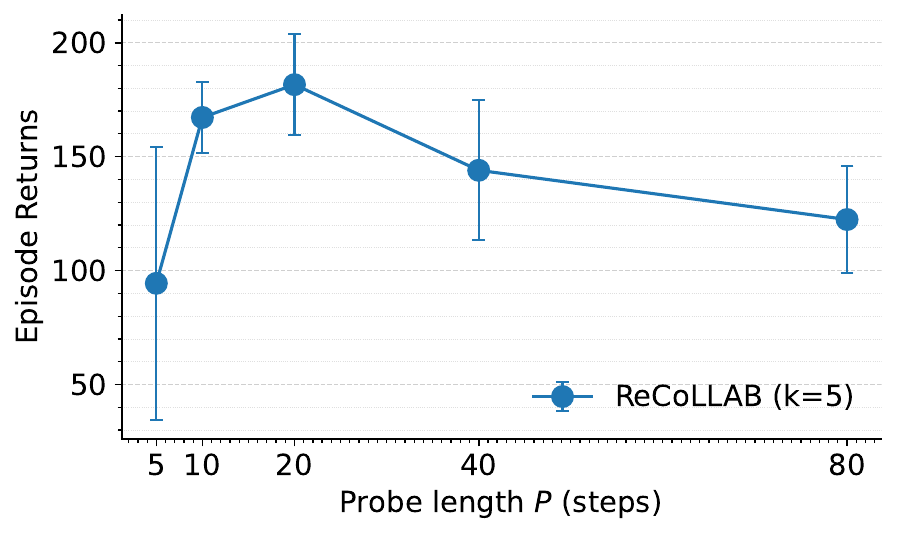}
        \caption{Episodic Returns vs. Probe-length}
        \label{fig:probe_sub2}
    \end{subfigure}
    \caption{\textbf{Probe-length ablations.} Teammate-type classification accuracy and episodic returns as a function of probe length $P$.}
    \label{fig:probe}
\end{figure}

\section{Ablation Studies}

In addition to the main results, we conduct ablation studies to analyze two key factors in teammate-type classification and adaptive performance: the length of the probe phase ($P$) and the number of retrieval exemplars ($k$). We focus on the \emph{Coordination Ring} layout which is the most difficult of the three in terms of teammate behavior complexity. These studies provide insight into the adaptation speed and robustness of each method.

\subsection{Effect of Probe Length \texorpdfstring{$P$}{P}}
Figure~\ref{fig:probe} reports classification accuracy and cumulative reward as a function of probe length $P \in \{5,10,20,40,80\}$. We find that both classification accuracy and cumulative returns are maximized at probe length $P=20$. Accuracy rises from near-random at $P=5$ to nearly perfect at $P=20$, after which performance begins to diminish. The dropoff in returns is due to longer probe phases increasing the time of sub-optimal cooperation which can be more difficult to break out of. These results highlight a trade-off between adaptation speed and performance: shorter probes allow faster decision-making but risk misclassification, while a probe at length $P=20$ strikes the best balance by achieving high accuracy and near-optimal returns without excessive delay.

\subsection{Effect of Retrieval Exemplars \texorpdfstring{$k$}{k}}
We next vary the number of retrieved exemplar trajectories $k \in \{1,3,5,10\}$ to evaluate the impact of retrieval grounding. Results are shown in Figure~\ref{fig:retrieval}. Varying the number of retrieved exemplars yielded only modest differences in both teammate-type classification accuracy
 and episodic returns. Accuracy remains relatively stable across values of $k$, and returns do not show consistent improvements with larger retrieval sets. This suggests that while retrieval grounding is important for \ReCollab’s overall performance, the precise number of exemplars plays a less critical role. In practice, small values such as $k=3$-$5$ appear sufficient to stabilize performance, and additional retrievals offer diminishing returns. These results highlight that the primary benefits of \ReCollab arise from conditioning on retrieved examples at all, rather than from scaling the retrieval set size.

\begin{figure}
    \centering
    \begin{subfigure}[b]{0.47\textwidth}
        \includegraphics[width=\linewidth]{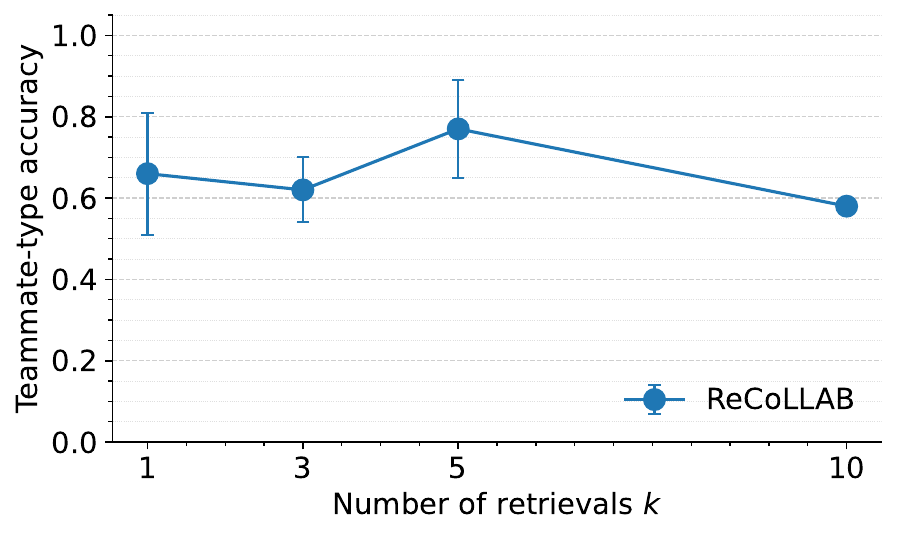}
        \caption{Classification Accuracy vs. Number of Retrievals}
        \label{fig:retrieval_sub1}
    \end{subfigure}\hfill
    \begin{subfigure}[b]{0.47\textwidth}
        \includegraphics[width=\linewidth]{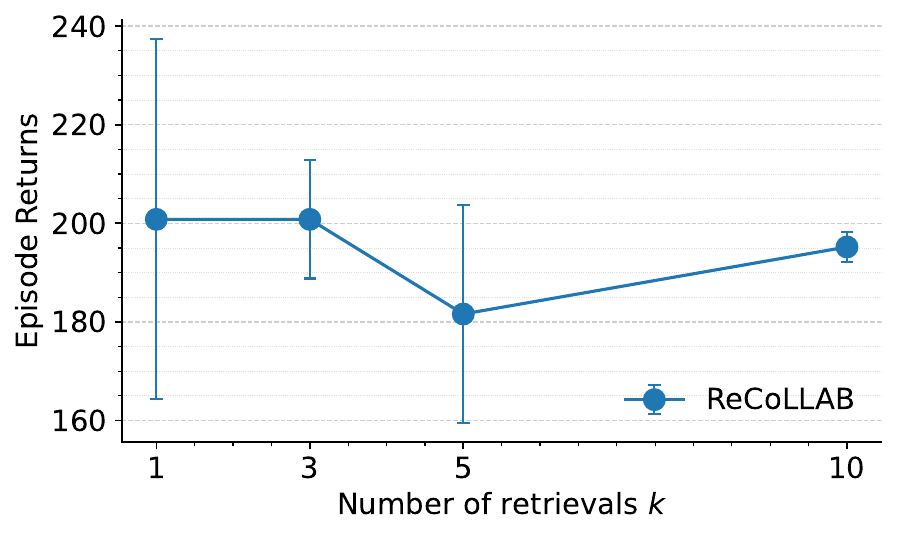}
        \caption{Episodic Returns vs. Number of Retrievals}
        \label{fig:retrieval_sub2}
    \end{subfigure}
    \caption{\textbf{Number of retrievals ablations.} Teammate-type classification accuracy and episodic returns as a function of the number of retrieved exemplars $k$.}
    \label{fig:retrieval}
\end{figure}

\section{Conclusion}
\label{sec:conclusion}

We introduced \textbf{\Collab}, an LLM-based framework for teammate type classification and policy routing in ad-hoc teamwork, and \textbf{\ReCollab}, its retrieval-augmented variant that grounds predictions in prior trajectory data.  
By framing type classification as a form of world modeling over teammates, our approach leverages the reasoning capabilities of LLMs while maintaining structured outputs for downstream control.

In the cooperative Overcooked domain, \ReCollab achieves competitive or superior early classification accuracy and team reward compared to the PLASTIC and Logistic Regression baselines. These findings suggest that LLM-based world models, when paired with targeted retrieval, can serve as effective agents in structured multi-agent RL settings without extensive fine-tuning. Our work opens several directions for future exploration, including scaling to continuous teammate behavior spaces, integrating online policy adaptation, and extending retrieval to multi-modal or human-agent teaming scenarios.

By bridging structured LLM reasoning, retrieval, and best-response policy selection, we take a step toward more flexible, generalizable agents capable of robust ad-hoc teamwork in complex environments.

\begin{ack}
This work was supported by the Office of Naval Research under Grant N000142412405 and the Army Research Office under Grants W911NF2110103 and W911NF2310363.
\end{ack}

\bibliographystyle{plainnat}
\bibliography{references}

\end{document}